\documentclass[12pt]{article}

\usepackage{latexsym}

\usepackage{graphicx}

\begin{document}


\title{Kaluza-Klein rotating multi-black hole configurations with electromagnetic field in Einstein-Maxwell-dilaton gravity}
\author{
     Stoytcho S. Yazadjiev \thanks{E-mail: yazad@phys.uni-sofia.bg}\\
{\footnotesize  Department of Theoretical Physics,
                Faculty of Physics, Sofia University,}\\
{\footnotesize  5 James Bourchier Boulevard, Sofia~1164, Bulgaria }\\
}

\date{}

\maketitle

\begin{abstract}
We present a new solution in 5D Einstein-Maxwell-dilaton gravity
describing an equilibrium configuration of extremal rotating black
holes with lens space horizon topologies. The basic properties of
the solution are investigated and the basic physical quantities are
calculated. It is shown that the black hole horizons are
superconducting in the sense that they expel the magnetic flux
lines.
\end{abstract}


\sloppy

In the last decade the higher dimensional gravity has attracted a
lot of interest and  it is now a well established area of the modern
theoretical and mathematical physics. The spacetimes with compact
extra dimensions (Kaluza-Klein spacetimes) take a special place
since they are much more realistic than the asymptotically flat
spacetimes. During the last years Kaluza-Klein black holes were
extensively studied and many exact solutions were found  but a lot
remains to be done in this direction since the spectrum of black
hole solutions in the Kaluza-Klein case is much richer than in the
asymptotically flat case. For recent review on the subject we refer
the reader to \cite{TI}.

In the present paper we deal with   multi-black hole Kaluza-Klein
spacetimes. Such multi-black hole configurations are very
interesting since they provide us with valuable insight into the
black hole theory in spacetimes with compact extra dimensions and
especially into the  black hole interactions in the strong field
regime. Very recently a Kaluza-Klein rotating vacuum multi-black
hole solution in five dimensions was studied in \cite{MIKT1}. The
authors show that the previously found solution of \cite{C} can be
interpreted as an equilibrium configuration of extremal rotating
black holes held apart by the repulsive spin-spin interaction. The
solution was generalized to the case of  5D Einstein-Maxwell gravity
and 5D minimal supergravity in \cite{MIKT2}.

The purpose of the present paper is to present  a new  solution
describing an equilibrium configuration of extremal rotating black
holes with selfgravitating electromagnetic field in the 5D
Einstein-Maxwell-dilaton gravity and to study some of its basic
properties. Our solution differs from those in \cite{MIKT2} in that
it is not charged and the magnetic fluxes through the horizons are
zero.


We consider the five dimensional Einstein-Maxwell-dilaton (EMD)
gravity given by the action

\begin{eqnarray}
S= \frac{1}{16\pi}\int d^4x\sqrt{-g} \left[{\cal R} -
2g^{\mu\nu}\partial_{\mu}\varphi\partial_{\nu}\varphi  -
e^{-2\alpha\varphi}F^{\mu\nu}F_{\mu\nu}\right]
\end{eqnarray}
where ${\cal R}$  is the Ricci scalar curvature with respect to the
spacetime metric $g_{\mu\nu}$, $F_{\mu\nu}$ is the Maxwell tensor,
$\varphi$ is the dilaton field and $\alpha$ is the dilaton coupling
constant. The field EMD equations are obtained by varying the
action:

\begin{eqnarray}
&&{\cal R}_{\mu\nu}= 2\nabla_{\mu}\varphi\nabla_{\nu}\varphi +
2e^{-2\alpha\varphi}\left[F_{\mu\sigma}F^{\sigma}_{\nu}-
\frac{g_{\mu\nu}}{6} F^{\rho\sigma}F_{\rho\sigma} \right],\\
&&\nabla_{\mu}\nabla^{\mu}\varphi = -
\frac{\alpha}{2}e^{-2\alpha\varphi} F^{\rho\sigma}F_{\rho\sigma},\\
&&\nabla_{\mu}\left[e^{-2\alpha\varphi}F^{\mu\nu}\right]=0.
\end{eqnarray}

In  the present paper we will consider the case $\alpha=\sqrt{8/3}$.
For this value of the dilaton coupling parameter we have found the
following solution

\begin{eqnarray}
&&ds^2 = V^{-1/3}\left\{-H^{-2}dt^2 + \left[\sqrt{2}(H^{-1}-1)dt +
Ld\psi + \sqrt{2}\cosh\gamma \, {\hat W}  \right]^2 \right\}
\nonumber \\ &&+ V^{2/3} H^2 \left[dx^2 + dy^2 + dz^2 \right], \\
&&A_{\mu}dx^{\mu}= -\frac{\sinh \gamma}{2}(H^{-1}-1)^2 dt -
\frac{\sinh\gamma}{\sqrt{2}}
(H^{-1}-1) Ld\psi - \cosh\gamma\sinh\gamma H^{-1} {\hat W}, \\
&&e^{\sqrt{2/3}\varphi}= V^{1/3} ,
\end{eqnarray}
where the metric function $V$ and the  1-form ${\hat W}$ are given
by

\begin{eqnarray}
&&V=\cosh^2\gamma - H^{-2}\sinh^2\gamma, \\
&&{\hat W} = \sum_{i} m_i \frac{z-z_i}{|R-R_{i}|} \frac{(x-x_{i})dy
- (y-y_{i})dx}{(x-x_i)^2 + (y-y_i)^2},
\end{eqnarray}
and $H$ is a harmonic function on the three dimensional flat space
explicitly given by

\begin{eqnarray}
H= 1 + \sum_i \frac{m_i}{|R-R_{i}|}
\end{eqnarray}
with point sources located  at $R_i=(x_i,y_i,z_i)$. The 1-form $\hat
W$ and the harmonic function $H$ satisfy the following equation on
the three dimensional flat space

\begin{eqnarray}
\nabla \times \hat W = \nabla H.
\end{eqnarray}

The parameters $\gamma$, $L$ $m_i$ run in the ranges
$-\infty<\gamma< \infty$, $L>0$ and $m_i\ge 0$. In the case
$\gamma=0$ our solution reduces to the vacuum  solution of
\cite{MIKT1}. In the general case the above metric possesses only
two Killing fields $\frac{\partial}{\partial t}$ and
$\frac{\partial}{\partial \psi}$.


Without loss of generality  we will  consider the case of two black
holes with $R_1=(0,0,0)$ and $R_2=(0,0,a)$ where $a>0$. In this case
the flat three metric, the harmonic  function $H$ and the 1-form
${\hat W}$ are given by

\begin{eqnarray}
&&ds^2_{3}= dx^2+ dy^2 + dz^2 = dR^2 + R^2\left(d\theta^2 + \sin^2\theta d\phi^2\right),\\
&&H= 1 + \frac{m_1}{R} + \frac{m_2}{\sqrt{R^2+ a^2 - 2Ra\cos\theta}},\\
&&{\hat W}= \left(m_1\cos\theta  + m_2 \frac{R\cos\theta -
a}{\sqrt{R^2 + a^2 - 2aR\cos\theta}}\right)d\phi.
\end{eqnarray}

In addition to the Killing fields $\frac{\partial}{\partial t}$ and
$\frac{\partial}{\partial \psi}$, in the case under consideration
there is one more Killing field $\frac{\partial}{\partial \phi}$.
The coordinates satisfy $-\infty<t<\infty$, $0<R<\infty$,
$0\le\theta\le\pi$, $0\le\phi<2\pi$ and $0\le \psi <2\pi$.

Using a standard approach   one can show that the point sources
$R=R_1$ and $R=R_2$ correspond to smooth horizons for the metric.
The coordinates we use to write solutions are actually singular on
the horizons and therefore we should introduce new coordinates that
cover the horizons too. We will do so only for the horizon at $R=0$
since   the other horizon can be treated analogously. Near the
horizon we introduce the new coordinates $\upsilon$ and $\tilde
\psi$ given by

\begin{eqnarray}
&& d\upsilon=dt - \cosh\gamma m_1 d\left(\frac{m_1}{R}\right), \\
&& d{\tilde \psi} = d\psi - \frac{\sqrt{2}}{L} dt -
\frac{\sqrt{2}\cosh\gamma m_1}{L} \frac{dR}{R} -
\frac{\sqrt{2}\cosh\gamma m_2}{L} d\phi.
\end{eqnarray}

In the new coordinates the metric takes the form

\begin{eqnarray}
ds^2 = \cosh^{-2/3}\gamma \left\{\frac{R^2}{m^2_1}d\upsilon^2 +
2\cosh\gamma d\upsilon dR  \right. \\ \left. + m_1^2 \cosh^2\gamma
\left[d\Omega^2 + 2 \left(\frac{L}{\sqrt{2}m_1\cosh\gamma}d{\tilde
\psi} + \cos\theta d\phi\right)^2 \right]  \right. \\ \left. +
4\cosh\gamma R \,d\upsilon \left(  \frac{L}{
\sqrt{2}m_1\cosh\gamma}d{\tilde \psi} + \cos\theta
d\phi\right)\right\}.
\end{eqnarray}

The spatial cross section of $R=0$  has the following induced metric

\begin{eqnarray}\label{LSM1}
dl^2=  m_1^2 \cosh^{4/3}\gamma \left[d\Omega^2 + 2
\left(\frac{L}{\sqrt{2}m_1\cosh\gamma}d{\tilde \psi} + \cos\theta
d\phi\right)^2 \right].
\end{eqnarray}
In order for the above metric to be smooth, just as in the vacuum
case \cite{MIKT1}, the following quantization condition must  be
imposed

\begin{eqnarray}
m_1= \frac{L}{2\sqrt{2}\cosh\gamma} n_1,
\end{eqnarray}
where $n_1$ is an integer.  With this condition imposed the metric
(\ref{LSM1}) describes the standard smooth  metric on the lens space
$L(n_1,1)$. In this way we obtained the analytical extension of our
metric across the surface $R=0$. In completely analogous way one can
build the analytical extension of the metric across the surface
$R_2$ with the following quantization condition

\begin{eqnarray}
m_2= \frac{L}{2\sqrt{2}\cosh\gamma} n_2,
\end{eqnarray}
where $n_2$ is an integer. Therefore the topology of the second
horizon is $L(n_2,1)$.

The asymptotic behavior of the solution is the following

\begin{eqnarray}
&&ds^2 \simeq \left(1 + \frac{2m\sinh^2\gamma}{R}\right)^{-1/3}
\left\{- \left(1- \frac{2m}{R}\right)dt^2 + \left(1 +
\frac{2m\cosh^2\gamma}{R}\right)\left(dR^2 + R^2d\Omega^2
\right)\right. \nonumber \\
&&\left. + \frac{n^2L^2}{4} \left(-\frac{2}{R\cosh\gamma} dt +
\frac{2d\psi}{n}+ \cos \theta
d\phi \right)^2 \right\},\\
&& A_{\mu}dx^{\mu}\simeq -\frac{\sinh\gamma m^2}{2 R^2} dt +
\frac{\sinh\gamma m}{\sqrt{2}R}L d\psi - \cosh\gamma \sinh\gamma m
\cos\theta d\phi,
\end{eqnarray}
with $m=\sum_i m_i$ and $n=\sum_i n_i$. From the explicit form of
the asymptotic metric it  is clear that the topology of spatial
infinity is $L(n,1)$.

It is worth noting that in the case of two black holes there is an
additional Killing vector $\frac{\partial}{\partial \phi}$ and the
techniques based on the notion of interval structure \cite{HY} can
be applied and they give the same results as the above analysis.

We proceed further with calculating the masses and angular momenta.
The mass of each black hole is given by the Komar integral

\begin{eqnarray}
M_{i}= \frac{-3}{32\pi} \int_{H_i} \star d\xi,
\end{eqnarray}
where $\star$ is the Hodge duality operator and $\xi$ is the 1-form
corresponding to the timelike Killing vector. The explicit
calculation gives the following result

\begin{eqnarray}
M_i= \frac{3}{2}\pi L m_{i}.
\end{eqnarray}

The total mass of the multi-black hole configuration is
\begin{eqnarray}\label{TM}
M= \frac{-3}{32\pi} \int_{\infty} \star d\xi= \frac{3}{2}\pi L
\sum_{i}m_i \left(1+ \frac{1}{3}\sinh^2\gamma\right).
\end{eqnarray}

The deference $M-\sum_i M_i \ne 0$  reflects the contribution of the
electromagnetic and dilaton fields to the total mass.

The angular momentum of each black hole is defined by the Komar
integral

\begin{eqnarray}
J^{\psi}_{i}= \frac{1}{16\pi}\int_{H_i} \star d\eta,
\end{eqnarray}
where $\eta$ is 1-form corresponding to the Killing field
$\frac{\partial}{\partial \psi}$. The explicit calculation gives

\begin{eqnarray}
J^{\psi}_{i}= \frac{L^2}{\sqrt{2}}\pi m_{i}.
\end{eqnarray}

The angular velocity of each black hole is
$\Omega_{H}=\frac{\sqrt{2}}{L}$ which, combined with the expressions
for the mass and angular momentum, results in the following relation

\begin{eqnarray}
M_{i}= \frac{3}{2}\Omega_{H}J^{\psi}_i.
\end{eqnarray}

For the total angular momentum of the configuration we find

\begin{eqnarray}
J^{\psi}= \frac{1}{16\pi}\int_{\infty} \star d\eta =
\frac{L^2}{\sqrt{2}}\pi \sum_im_i= \sum_i J^{\psi}_i.
\end{eqnarray}
Therefore the electromagnetic field gives no contribution to the
total angular momentum.

The total electric charge of the configuration is defined by

\begin{eqnarray}
Q= \frac{1}{4\pi} \int_{\Sigma}e^{-2\alpha\varphi}\star F
\end{eqnarray}
and the calculations give $Q=0$. This result can be easily seen from
the asymptotic behavior  $A_t\sim \frac{1}{R^2}$. The electric
charge of each black hole in the configuration is also zero,
$Q_i=0$. The nontrivial quantity that characterizes the
electromagnetic field is the magnetic flux $\Gamma$ through the base
space $S_{\infty}^2$  of the $S^1$-fibration $L(n,1)$ at infinity,
namely

\begin{eqnarray}
\Gamma= \frac{1}{4\pi} \int_{S^2_{\infty}}F= \cosh\gamma\sinh\gamma
m.
\end{eqnarray}

It is also interesting  to find the magnetic flux through a portion
of the horizon cross sections. It is known however, that in the
general case, the magnetic flux lines are expelled from the extremal
black holes, more precisely the components of the field strength
normal to the horizon vanish \cite{BD},\cite{CEG}. In our case we
have

\begin{eqnarray}
F|_{H_i}=0
\end{eqnarray}
and therefore, within the framework of the present solution, the
black hole horizons are superconducting in the sense that they
exhibit "Meissner effect" typical for the superconductors. Even
more, the electric field also vanishes on the horizons.

We now are at a position to derive a Smarr-like formula giving a
relation between the Komar mass and the magnetic flux. For this
purpose we consider the 1-form $\chi=i_{\eta} i_{\xi}
e^{-2\alpha\varphi} \star F $ which is closed  $d \left( i_{\eta}
i_{\xi} e^{-2\alpha\varphi} \star F \right)=0$ as a consequence of
the Killing symmetries and the Maxwell equations. The 1-form $\chi$
is invariant under the Killing symmetries and therefore it can be
viewed as defined on the orbit (factor) space {\it
spacetime/isometry group}. Since the factor space is simply
connected \cite{HY}, there exists a potential $\Psi$ such that
$d\Psi=i_{\eta} i_{\xi} e^{-2\alpha\varphi} \star F$. The explicit
expression for $\Psi$ in our case is

\begin{eqnarray}
\Psi= - \frac{\cosh\gamma\sinh\gamma}{2} \frac{1-
H^{-2}}{\cosh^2\gamma - H^{-2}\sinh^2\gamma} +
\frac{\tanh\gamma}{2},
\end{eqnarray}
where we have fixed the arbitrary constant in the definition of
$\Psi$ so that $\Psi$ vanishes on the horizons.  Using then the same
approach as in \cite{NY} it can be shown that the following relation
is satisfied

\begin{eqnarray}
M=\sum_i M_i + \pi L \Psi(\infty) \Gamma.
\end{eqnarray}
This is in fact eq.(\ref{TM}) as can be checked.

The ergosurface, defined by $g(\xi,\xi)=0$, exists always since
$g(\xi,\xi)|_{H_i}>0$ and $g(\xi,\xi)|_{\infty}<0$ and its topology
depends on the point sources configuration just as in the vacuum
case.

As a final remark it is worth noting that our preliminary
investigations show that there exist  solutions describing rotating
extremal multi-black hole configurations with selfgravitating
electromagnetic field, more general than the solutions presented in
\cite{MIKT2} and in the present paper. The results will be presented
elsewhere.


\section*{Acknowledgements}
The partial financial supports from the  Bulgarian National Science
Fund under Grant DMU-03/6, and by Sofia University Research Fund
under Grant 148/2012 are gratefully acknowledged.

\end{document}